\documentclass[11pt]{article}
\pdfoutput=1
\usepackage{jheppub}
\usepackage{amssymb,amsfonts}
\usepackage{mathrsfs}
\usepackage{slashed}
\usepackage{graphics}
\usepackage{tikz}
\usetikzlibrary{decorations.markings}
\let\savenumberline\numberline
\def\numberline#1{\savenumberline{#1.}}
\makeatletter
\renewcommand{\@seccntformat}[1]{\csname the#1\endcsname.\,\,}
\makeatother
\def\Bucket{\rotatebox[origin=c]{180}{\(\Omega\)}}

\renewcommand{\tilde}[1]{\widetilde{#1}}
\renewcommand{\hat}[1]{\widehat{#1}}
\newcommand{\be}{\begin{equation}}
\newcommand{\ee}{\end{equation}}
\newcommand{\bea}{\begin{eqnarray}}
\newcommand{\eea}{\end{eqnarray}}

\newcommand\secref[1]{{\S\ref{#1}}}

\newcommand\figref[1]{{Figure~\ref{#1}}}

\makeatletter
\def\@fpheader{\relax}
\makeatother
\usepackage{graphicx}
\usepackage{latexsym}

\title{\ \vspace{1.5in} \\ \hbox{Tropical Limits of Dirac Operators}}
\author{Emil Albrychiewicz, Andr\'{e}s Franco Valiente and Vi Hong}
\affiliation{\medskip
Leinweber Institute for Theoretical Physics and Department of Physics\\
University of California, Berkeley, CA, 94720-7300, USA\medskip\\
Theoretical Physics Group, Lawrence Berkeley National Laboratory\\
Berkeley, CA 94720-8162, USA}
\emailAdd{ealbrych@berkeley.edu}
\emailAdd{andresfranco@berkeley.edu}
\emailAdd{vihong14@berkeley.edu}
\abstract{We explore the tropical analog of spinors by representing tropical geometries as foliated Riemann surfaces endowed with degenerate complex structures. We investigate tropical limits of the Laplace-Beltrami operator and explicitly construct its square root, which defines a tropical Dirac operator. We find that the tropical Clifford algebra is classified as a degenerate Clifford algebra with nilpotent generators. The nilpotent generator allows us to work with a new kind of representation that allows for Grassmann odd numbers, effectively supersymmetrizing the tropical spin bundle. We show through Dirac-Bergmann's quantization procedure, that the corresponding tropicalized quantum field theories enjoy a purely fermionic topological symmetry which can be expected to give a new class of path integral localization that we call tropical localization similar to the alternative localization method recently constructed by Choi and Takhtajan. We also discuss how the tropical Dirac operator, when twisted by gauge fields, obeys a tropical version of the Lichnerowicz identity, thereby demonstrating how some elements of Yang-Mills curvature should arise in the tropical limit.
}
\begin{document}
\maketitle

\section{Introduction}
In \cite{trsm}, it was demonstrated that tropical geometry \cite{msintro, rau, mikhalkinrau} can be investigated through the Maslov dequantization limit \cite{litvinov, viro, virohyper} of topological sigma models \cite{ewtsm}, leading to the introduction of \textit{tropological sigma models}. These tropological sigma models possess a worldsheet formulation that exhibits a non-relativistic foliation structure, with the target space naturally corresponding to the tropical geometry under consideration. The emergence of these foliations is attributed to the degeneration of the complex structures in the original sigma models, which results in nilpotent endomorphisms of the tangent bundle known as Jordan structures. Consequently, the path integral formulation of tropological sigma models provides an alternative to working directly with real algebraic varieties, which typically appear in tropical geometry. Instead, this approach offers a more physics-friendly representation, where tropical geometries manifest as foliated Riemann surfaces. A key result is that the correlation functions of the original topological sigma model coincide with those of its tropicalized counterpart, offering a novel and computationally efficient method through the lens of tropical combinatorics for evaluating these objects.

It was further suggested in \cite{trsm}, that one should be able to investigate topological invariants that are naturally associated to odd-dimensional geometries by extending the Jordan structures of foliated complex geometries to e.g., contact manifolds. In particular, the Maslov dequantization limit remains well-defined in odd-dimensional geometries, providing a natural path for such an extension. In order to extract interesting information about potential tropical contact invariants, one may explore tropicalized index theorems associated with differential operators constructed on these foliated geometries. However, an immediate obstacle in this pursuit is formulating a notion of tropical supersymmetry for foliated worldsheet theories and their associated tropical target spaces which is needed for functional integral localization. Thus, as a first step toward formulating tropical supersymmetry, it is necessary to develop a theory of tropical spinors. We use the results developed here in an upcoming work \cite{WIP2} on supersymmetric worldline theories in the tropical limit.

In this paper, we establish some standard conventions for Riemann surfaces in section \secref{sec:ComplexAndTropical} and their tropicalization in the sense of \cite{trsm}. We investigate naive tropical limits of the Laplace-Beltrami operator and give distinct definitions for tropical harmonic functions that lead to piecewise linear curves. We see that one of these limits gives a linear operator, which admits easily constructable square roots for the case of foliated Riemann surfaces $\Sigma$, while the other limit gives a nonlinear operator related to the Hamilton-Jacobi equation. We call the former operator, a \textit{tropicalized Laplacian}. 

In section \secref{sec:TropSpin}, we extend the 2-dimensional tropicalized Laplacian to act on a 3-dimensional manifold that can be decomposed as a product of a Riemann surface $\Sigma$ and real line $\mathbb{R}$ in order to more clearly see the non-trivial algebraic structure of the square root of the tropical Laplacian. We find that the square root of the tropical 3-dimensional Laplacian gives rise to an 8-dimensional algebra that can be classified as a degenerate Clifford algebra which admits a nilpotent generator associated to the null direction that we tropicalize. In order to construct complex irreducible representations for the tropical Clifford algebra, one is forced to supersymmetrize the spinor bundle \cite{Clifford} in order to take into account the nilpotent generator. The associated square root is a differential operator invariant under foliation preserving diffeomorphisms which we call a \textit{tropicalized Dirac operator}. We demonstrate that there are two inequivalent degenerations of Riemann surfaces; one that induces a tropical notion of spinor and another which produces a counterpart that we call an \textit{arctic spinor}\footnote{Here, the  arctic limit refers to the singular limit that is the opposite limit of the tropical one.} which arises from the fact there are two choices for which direction the foliations can run along.  

In section \secref{sec:PropTropDir}, we study the solutions of the tropicalized Dirac operator, and construct twisted tropicalized Dirac operators. From the Lichnerowicz identity \cite{lichnerowicz1963geometrie} of the twisted Dirac operator, one can see how tropical versions of Yang-Mills curvatures \cite{Yang:1954ek} can affect the dynamics of tropical spinor fields.  We discuss the Dirac-Bergmann \cite{Bergmann:1949zz, Dirac:1950pj, Anderson:1951ta, Henneaux:1992ig} quantization associated to tropical spinor fields and discuss the essentials in constructing a novel type of path integral that takes into account the Grassmann number introduced by the the nilpotency of the degenerate directions.  In section \secref{sec:TropLoc}, we discuss how the consequences of having of having a Grassmann number in the spinor bundle should result in a new kind of fermionic localization similar to what was similarly found in the context of worldline localization \cite{Choi:2025jis}. As an example, we use this approach to calculate Schwinger pair production rate \cite{Schwinger:1951nm} and recover the original result \cite{Affleck:1981bma}. We leave some open questions that can lead to fruitful followups, such as the formulation of tropical supersymmetry and the path integral quantization of these tropical spinors.

\section{Complex and Tropical Geometry}
\label{sec:ComplexAndTropical}

In this section, we review the extraction process of tropical geometry from complex geometry through the Maslov dequantization limit. We discuss the degenerations of common geometric structures that are available on Riemann surfaces as well as relevant differential operators. We begin with a Riemann surface $\hat\Sigma$ that admits local complex coordinates $(z,\bar{z})$. For clarity, we put a hat on geometric structures and spaces that have not yet undergone the tropical limit. We remove the hat once they are completely deformed into their tropical counterpart. We do not apply this convention to coordinates since the tropical coordinates are denoted as $(r,\theta)$. 

The Maslov dequantization limit that leads to tropical geometries is implemented by deforming the local coordinates in the following way
\begin{equation}
\label{eqn:SubTropDef}
z=e^{\frac{r}{\hbar}+i \theta}, \quad \bar{z}=e^{\frac{r}{\hbar}-i \theta}.
\end{equation}
By the tensor transformation law, the derivatives then take the following form
\begin{equation}
\begin{aligned}
\partial_z & =e^{-\left(\frac{r}{\hbar}+i \theta\right)}\left(\frac{\hbar}{2} \partial_r-\frac{i}{2} \partial_\theta\right), \\
\partial_{\bar{z}} & =e^{-\left(\frac{r}{\hbar}-i \theta\right)}\left(\frac{\hbar}{2} \partial_r+\frac{i}{2} \partial_\theta\right).
\end{aligned}
\end{equation}
Notice that we have a convergent limit as $\hbar \rightarrow 0 $ for the derivative operators. The abelian differentials are also deformed as
\begin{equation}
\begin{aligned}
\label{eqn:DzTropical}
& d z=e^{\frac{r}{\hbar}+i \theta}\left(\frac{1}{\hbar} d r+i d \theta\right), \\
& d \bar{z}=e^{\frac{r}{\hbar}-i \theta}\left(\frac{1}{\hbar} d r-i d \theta\right).
\end{aligned}
\end{equation}
As discussed in \cite{trsm}, for $\hbar\neq 0$, this is simply a change of coordinates and hence, strictly speaking, does not deform the geometry yet; nonetheless, we generically call this the subtropical deformation. It is only in the singular limit $\hbar\rightarrow 0$, that we recover tropical geometry. 

Our Riemann surfaces is equipped with an almost complex structure $\hat{\varepsilon}$ and a metric tensor $\hat{g}$. We denote the inverse metric tensor as $\hat{h}$. We can choose local complex coordinates such that the line element takes the following form
\begin{equation}
\hat{ds^2}=g_{z \bar{z}} d z d \bar{z}.
\end{equation}
The subtropical deformation yields
\begin{equation}
\hat{ds^2}=g_{z \bar{z}} e^{\frac{2 r}{\hbar}}\left(\frac{1}{\hbar^2} d r^2+d \theta^2\right).
\end{equation}
With a multiplicative renormalization, the tropical line element in the limit $\hbar\rightarrow 0$ reduces to
\begin{equation}
    ds^2= dr^2,
\end{equation}
from which we can extract the components of the metric tensor. We use $\doteq$ to denote the matrix representation constructed using the adapted local coordinates
\begin{equation}
    g \doteq \left[\begin{array}{cc}
1 & 0 \\
0 &0
\end{array}\right].
\end{equation}

Taking the same subtropical deformation of the inverse metric tensor $\hat{h}$ in the limit $\hbar\rightarrow 0$ leads to
\begin{equation}
    h \doteq \left[\begin{array}{cc}
0 & 0 \\
0 & 1
\end{array}\right].
\end{equation}
We observe that the tropical limit of the inverse metric tensor is not the inverse of the tropical metric tensor; instead, they satisfy a mutual invisibility law of the form
\begin{equation}
    hg=gh\doteq \left[\begin{array}{cc}
0 & 0 \\
0 & 0
\end{array}\right].
\end{equation}

Before taking the tropical limit, it is well known that one may represent a conformal class of metrics equivalently by a complex structure characterized by $\hat{\varepsilon}^2=-\text{Id}$. Since a complex structure may be represented as an endomorphism of the tangent bundle, one may compute the Maslov dequantization and after a renormalization, one obtains
\begin{equation}
\varepsilon=\left[\begin{array}{cc}
0 & 1 \\
0 & 0
\end{array}\right].
\end{equation}
This tropicalized complex structure is a nilpotent endomorphism of the tangent bundle. Its matrix representation is reminiscent of the Jordan normal form of matrices and hence, it is known as a Jordan structure. 

This Jordan structure on the tropicalized two-dimensional manifold $\Sigma$, creates a fiber-wise filtration on the tangent bundle, which can be used to define an integrable distribution which then induces a natural foliation on the surface. The filtration structure naturally extends to the full tensor algebra over the tangent space. Hence, instead of working directly in the quotient space where tropical geometries are represented by 1 real-dimensional  algebraic varieties, we instead represent the tropical geometry by a 1 complex-dimensional foliated Riemann surface.

One finds that the nilpotency of the Jordan structure is preserved under foliation-preserving diffeomorphisms
\begin{equation}
\begin{aligned}
& \widetilde{r}=\widetilde{r}(r), \\
& \widetilde{\theta}=\widetilde{\theta}_0(r)+\theta \partial_r \widetilde{r}(r).
\end{aligned}
\end{equation}
By looking at the Lie algebra associated to this symmetry group, we obtain
\begin{equation}
\begin{aligned}
\delta r & =f(r), \\
\delta \theta & =F(r)+\theta \partial_r f(r).
\end{aligned}
\end{equation}
Thus, the local symmetries of the Jordan structure are generated by the infinite-dimensional Lie algebra whose elements are parametrized by two real, arbitrary, projectable, and differentiable functions $f(r)$ and $F(r)$ on the foliation. Projectability implies that it is leafwise constant, i.e. a basic function.

One can also take the opposite Maslov dequantization limit such that $\hbar\rightarrow \infty $. In this case, we generally cannot interpret the resulting geometry as a tropical geometry. Instead, we have a different kind of geometry that we call an \textit{arctic geometry}\footnote{We want to give credit to Hořava's family for suggesting the name, ``arcticlization" however we want to note that there have been previous usage of the term ``arctic semrings" in the mathematical literature \cite{Perrin1992}}. The arctic geometry limit of a Riemann surface can still be represented by a foliated complex manifold. However, the foliation is now tangent to the other coordinate direction $r$. For example, we can construct the subarctic deformation of geometric objects in the analogical way. as in \eqref{eqn:SubTropDef}, by using the coordinates 
\begin{equation}
z=e^{\hbar' r+i \theta}, \quad \bar{z}=e^{\hbar' r-i \theta}.
\end{equation}
Using this redefined Maslov dequantization, we can now send $\hbar'\rightarrow 0$ to obtain the arctic geometry. For the rest of this section, we focus our attention on the tropical deformations.

Prior to taking the tropical limit, it is usually quite fruitful to study a natural class of differential operators on the manifold, in order to extract interesting geometric-analytic and topological data. The most natural candidate would be the exterior derivative. However, it is important to note that even though Maslov dequantization acts on coordinates, we wish to preserve the  differential structure of the Riemann surface $\Sigma$ and hence we do not deform objects like the exterior derivative. As an example, one might be tempted to construct a subtropical exterior derivative
\begin{align*}
    \hat{d} f=\frac{1}{\hbar} \partial_r f d r+\partial_\theta f d \theta.
\end{align*} 
After a rescaling and taking the tropical limit, one would obtain
\begin{equation}
d f= \partial_r f (r,\theta)  d r.
\end{equation}
However, this would be inconsistent with the current treatment of the differential structure in the tropological sigma model construction \cite{trsm}. We expect one-forms and sections of all tensor algebras to have information about the direction tangential to the leaves of the foliation $d\theta$ and transverse to the foliation $dr$.  A consequence of this interpretation is that two independent cohomology theories may be developed. The first is the canonical tangential/leafwise cohomology theory that can be obtained by restricting the deRham cohomology to the space of leaves. This is in contrast to the naive subtropical limit of the exterior derivative which suggests a cohomology theory that is transverse to the leaves. In order to capture the topology transverse to the leaves of the foliation, one might instead look at the Haefliger cohomology \cite{haefliger1979differential} which can be used to construct natural characteristic classes on the foliation.

For the simple case of a non-singular foliated Riemann surface, one is able to construct a foliation of codimension one by looking at the level surfaces defined by a real-valued function $f$ on the foliation without any critical points. Its differential, $\omega=df$, is therefore locally nonsingular. To ensure that this differential exists globally,  the kernel $\operatorname{Ker}\omega$ must define an integrable distribution which then gives us the foliation of interest. One can show that such an integrability condition is given by $\omega \wedge d \omega=0$, which is trivially zero for a Riemann surface, effectively recovering the statement of the Frobenius theorem for integrable distributions on surfaces. This construction can be extended to higher dimensional foliated geometries and gives rise to the simplest characteristic classes on foliations such as the Godbillion-Vey \cite{godbillon1971invariant, thurston1972noncobordant, tondeur1997geometry}, which is a degree 3 form which measures how far away our distribution is from being integrable. Some preliminary constructions of how the tropical limit should be taken in higher dimensions were given in \cite{Albrychiewicz:2025afk}. The main result was that the tropical geometries are generically associated to filtered manifolds instead of simply foliated manifolds. We leave the question of what sort of natural cohomology theory should be used in higher dimensions as an open research direction.

Our foliated Riemann surfaces allow for the possibility of singular foliations that represent the vertex of the underlying tropical graph. Consequently, the topology of the foliation can become much more complicated but should still satisfy a global balancing condition at the singular point. In this particular case, it is possible that the singularity of the foliation can yield secondary characteristic classes which are local invariants that measure the complexity of the singular point. We will not consider these potential issues for the rest of the paper since we are interested in the construction of a tropical limit of Dirac operator on the edge of a tropical graph before we reach the vertex, or equivalently, an infinitely long foliated cylinder. 

Instead of considering the exterior derivative, we consider differential operators that depend on the geometric data deformed under the tropical limit. In particular, if we equip our Riemann surface $\hat{\Sigma}$ with a metric tensor $\hat{g}$, then we can consider the Laplace-Beltrami operator and it's natural generalizations to the exterior algebra, the Hodge-DeRham Laplacian.

Since we are interested in the explicit construction of a tropical Dirac operator on a non-singular foliated Euclidean space, we avoid topological obstructions since Euclidean space is topologically trivial. Before we try to investigate what a tropical Dirac equation looks like, we would like to investigate what sort of structure a tropical Laplacian has.  Given that we have a nondegenerate metric tensor, one can construct a diffeomorphism invariant inner product on the space of smooth functions. The inner product of any two functions $f_1, f_2 $ on $\hat\Sigma$ is given as
\begin{equation}
\left\langle f_1 \mid f_2\right\rangle=\int_{\hat\Sigma} d^2 \sigma \sqrt{\hat{g}} f_1(\sigma) f_2(\sigma).
\end{equation}
For 1-forms $\omega_1,\omega_2$, the diffeomorphism invariant inner product is then
\begin{equation}
\left\langle\omega_1 \mid \omega_2\right\rangle=\int_{\hat\Sigma} d^2 \sigma \sqrt{\hat{g}} \hat{g}^{\alpha \beta}(\sigma) \omega_\alpha (\sigma)\omega_\beta (\sigma).
\end{equation}
Using these inner products, a natural adjoint operator $\hat{d}^{\dagger}$ can then be constructed. This can then be used to define the Laplace-Beltrami operator on the algebra of functions and extended linearly to the Hodge-Laplacian on the exterior algebra $\bigwedge T^*\Sigma$.  In general, the Laplacian is defined by
\begin{equation}
\hat{\Delta}=\hat{d}^{\dagger}\, \hat{d}+\hat{d}\, \hat{d}^{\dagger}.
\end{equation}
In local coordinates, the Laplace-Beltrami operator takes the form
\begin{equation}
\hat{\Delta}f=\frac{1}{\sqrt{\hat{g}}} \partial_i\left(\sqrt{\hat{g}}\, \hat{g}^{i j} \partial_j f \right).
\end{equation}
In the case of a flat Riemann surface, the Laplace-Beltrami operator reduces down to
\begin{equation}
\label{eqn:LaplaceBeltrami}
\hat{\Delta} f=4 \partial_z \partial_{\bar{z}} f.
\end{equation}
Then, we would like to see if this structure can be written as the square of a traditional operator or if one needs to refine what one means by a Clifford algebra for tropical geometries such that tropical Laplacians admit a square root. 

The subtropical deformation of the Laplace-Beltrami operator \eqref{eqn:LaplaceBeltrami} takes the following form
\begin{equation}
\hat{\Delta} f=e^{-\frac{2 r}{\hbar}}\left[\hbar^2 \partial_r^2 f+\partial_\theta^2 f\right].
\end{equation}
In the asymptotic limit $\hbar\rightarrow 0$, this reduces down to
\begin{equation}
\Delta f = e ^{-\frac{2 r}{\hbar}} \partial^2_{\theta} f.
\end{equation}
In order to extract a finite differential operator, we can regulate this limit by rescaling the original Laplacian such that the tropicalized Laplacian is defined as
\begin{equation}
\label{eqn:TropLap}
\Delta=\lim _{\hbar \rightarrow 0} e^{\frac{2 r}{\hbar}} \hat{\Delta}=\partial_\theta^2.
\end{equation}
This rescaling can be motivated by the fact that the volume form $dz\wedge d\bar{z}$ on a Riemann surface picks up an additional factor of $e^{\frac{2r}{\hbar}}$ upon Maslov dequantization (cf. \eqref{eqn:DzTropical}). 

A natural guess for the square root of the operator \eqref{eqn:TropLap} is simply $\partial_{\theta}$. However, this is in conflict with the usual construction of the Dirac operator being generated by a Clifford deformation of the exterior algebra, which should know about all directions on the manifold. We present the correct construction in the next section \secref{sec:TropSpin} and find that the Dirac operator still takes the following form in the adapted coordinates
\begin{equation}
    \slashed{D}=\gamma^r\partial_r+\gamma^\theta \partial_\theta.
\end{equation}
Here, the tropical Clifford elements $\gamma^r, \gamma^\theta$ satisfy a different algebraic structure in contrast to the standard case where we have a non-degenerate Clifford algebra. In fact, one can show that the algebraic structure of the tropicalized Clifford is isomorphic to a degenerate Clifford algebra, which we will identify in \secref{sec:TropSpin}.

Tropical harmonic functions are then defined relative to $\Delta$ as $\Delta f=0$; the local solutions have the form
\begin{equation}
f(r, \theta)=f_0(r)+\theta f_1(r),
\end{equation}
with arbitrary, smooth, and projectable functions $f_0(r)$ and $f_1(r)$. Imposing the periodicity of $\theta\sim \theta+2\pi$ forces $f_1(r)$ to vanish, hence, we do not seem to recover a non-trivial tropical curve when $f$ is a real-valued function however if we allow $f$ to take values in $S^1$, then we recover a winding mode and can write this as
\begin{equation}
f(r, \theta)=f_0(r)+n\theta,
\end{equation}
for $n \in \mathbb{Z}$. In this particular case, we can then explicitly write down the balancing condition for circle-valued functions by gluing together multiple local patches.

Alternatively, we may deform an arbitrary smooth test function $f(r,\theta)$ in a well-chosen manner.  Since this is a singular limit, many inequivalent limits can be taken by choosing how one deforms the algebra of smooth functions.  In \cite{trsm}, it was shown that allowing the following deformation of the algebra of functions,

\begin{equation}
f=\exp \left(\frac{F(r,\theta)}{\hbar}+i \Theta(r,\theta)\right),
\end{equation}
 leads to interesting results. For example, in the case of topological sigma models, deforming the fields in this particular way yields an alternative calculation for Gromov-Witten invariants. In this case, we end up with the following expression 
\begin{equation}
\begin{aligned}
\frac{\hat{\Delta} f}{f} & =e^{-\frac{2 r}{\hbar}}\left[\left(\partial_r F\right)^2-\left(\partial_\theta \Theta\right)^2+\frac{1}{\hbar^2}\left(\partial_\theta F\right)^2-\hbar^2\left(\partial_r \Theta\right)^2+\hbar \partial_r^2 F+\frac{1}{\hbar} \partial_\theta^2 F\right] \\
& +i e^{-\frac{2 r}{\hbar}}\left[2 \hbar \partial_r F \partial_r \Theta+\frac{2 \partial_\theta F \partial_\theta \Theta}{\hbar}+\hbar^2 \partial_r^2 \Theta+\partial_\theta^2 \Theta\right].
\end{aligned}
\end{equation}
After a multiplicative rescaling, taking the naive limit suggests that
\begin{equation}
\begin{aligned}
\label{eqn:FirstDef}
\bar{\Delta}f=\lim _{\hbar \rightarrow 0} e^{\frac{2r}{\hbar}}\frac{\hat{\Delta} f}{f} & =\left(\partial_\theta F\right)^2+i (2 \partial_\theta F \partial_\theta \Theta).
\end{aligned}
\end{equation}
This limit doesn't yield any useful differential operator but it provides a hint as to what class of functions we should consider. In particular, if we restrict to projectable radial functions i.e.,  $\partial_\theta F=0$, we can take an alternative sensible limit. Thus, the subtropical Laplacian becomes
\begin{equation}
\begin{aligned}
\frac{\hat{\Delta} f}{f} & =e^{-\frac{2 r}{\hbar}}\left[\left(\partial_r F\right)^2-\left(\partial_\theta \Theta\right)^2-\hbar^2\left(\partial_r \Theta\right)^2+\hbar \partial_r^2 F\right] \\
& +i e^{-\frac{2 r}{\hbar}}\left[2 \hbar \partial_r F \partial_r \Theta+\hbar^2 \partial_r^2 \Theta+\partial_\theta^2 \Theta\right].
\end{aligned}
\end{equation}
We can extract a convergent expression by defining an alternative tropical Laplacian $\tilde{\Delta}$, c.f. \eqref{eqn:TropLap}, as follows
\begin{equation}
\label{eqn:TropAltLap}
\tilde{\Delta} f=\lim _{\hbar \rightarrow 0} e^{\frac{2 r}{\hbar}} \frac{\hat{\Delta} f}{f}= \left[\left(\partial_r F\right)^2-\left(\partial_\theta \Theta\right)^2\right] 
+i \left[\partial_\theta^2 \Theta\right].
\end{equation}
In the tropical limit, we recover an analytically continued Jacobi-Hamilton equation with an additional constraint. Unlike the first definition for tropical Laplacian $\Delta$ \eqref{eqn:FirstDef} that we proposed, this definition for a tropical Laplacian $\tilde{{\Delta}}$ is nonlinear, and hence constructing a Dirac operator for it appears to be more difficult. Tilde tropical harmonic functions are then defined as projectable functions that satisfy real and imaginary part of \eqref{eqn:TropAltLap}
\begin{equation}
\begin{gathered}
\left(\partial_r F\right)^2-\left(\partial_\theta \Theta\right)^2=0, \\
\partial_\theta^2 \Theta=0.
\end{gathered}
\end{equation}
Solving these equations, it can be shown that tilde tropical harmonic functions are now locally characterized by affine curves
\begin{equation}
\begin{aligned}
& \Theta(r,\theta)=\Theta_0(r)+\theta  \Theta_1(r), \\
& F(r)=F_0+\int^r_0 d\tilde{r}\,\Theta_1(\tilde{r}),
\end{aligned}
\end{equation}
with $F_0$ being a constant.

Recall that the coordinate $\theta$ and the tropical phase function $\Theta$ are both defined up to a $2\pi$ periodicity. This gives us the final expression for tilde tropical harmonic functions
\begin{equation}
\begin{aligned}
& \Theta(r, \theta)=\Theta_0(r)+n \theta, \\
& F(r)=F_0+n r.
\end{aligned}
\end{equation}
In these equations, $n$ is again an integer that can be interpreted as the winding number associated to the periodicity of $\Theta$. We now see that for this definition of tropical harmonic functions, the tilde tropical harmonic functions are locally characterized by affine curves on a foliated Riemann surface whose tropical phase function is a genuine tropical curve  i.e., affine curves with integer valued coefficients up to a $\theta$ dependent shift. Comparing both proposed definitions for the tropical Laplacian suggests that we can either have a simple linear differential operator $\Delta$ \eqref{eqn:TropLap} with admit tropical curves for circle valued functions, and additionally admits a tropical Dirac operator $D$ given by a degenerate Clifford algebra,  or have a complicated nonlinear differential operator $\tilde{{\Delta}}$ \eqref{eqn:TropAltLap} that also leads to non-trivial tropical curves however its corresponding square root is not obvious. For the rest of this paper, we focus on $\Delta$ since the construction of its square root is straightforward.

We want to emphasize that the Maslov dequantization procedure makes sense for odd-dimensional geometries as well and hence be used to construct generalized tropical geometries in terms of foliated manifolds. We outline how to do this for the simplest setting in the next section.

\section{Tropical Spin Geometry}
\label{sec:TropSpin}

It is well known that the topology of a space can bring additional constraints and obstructions in the construction of Dirac operators. As mentioned before, we temporarily avoid this issue by trying to construct local Dirac operators on a 3-dimensional Euclidean space. Hence, we extend our Riemann surface $\hat{\Sigma}$ to a 3-dimensional manifold $\hat{N}$ which can be decomposed as a trivial product $\hat{\Sigma} \times \mathbb{R}$. For simplicity, we start with the case of a flat Riemann surface. We place local coordinates $(x,y,\tau)$ on $\hat{N}$ and equip it with a Euclidean metric tensor whose line element is $\hat{ds^2}=dx^2+dy^2+d\tau^2$.  Switching $(x,y)$ to complex coordinates $(z,\bar z)$ on $\Sigma$, performing the subtropical deformation on $\Sigma$, and also performing an anisotropic Weyl scaling of the metric in the sense of \cite{trsm} so that $(z,\bar{z})$ coordinates has the same scaling  as the $\tau$ coordinate, the line element now takes the form
\begin{equation}
\hat{d s^2}=\left(\frac{d r^2}{\hbar^2}+d \theta^2+d \tau^2\right).
\end{equation}
The corresponding subtropical Laplace-Beltrami operator is then
\begin{equation}
\hat{\Delta} =\left(\hbar^2 \frac{\partial^2 }{\partial r^2}+\frac{\partial^2 }{\partial \theta^2} +\frac{\partial^2 }{\partial \tau^2}\right).
\end{equation}
Applying Maslov dequantization, as outlined in section \secref{sec:ComplexAndTropical}, we obtain a convergent limit that gives the tropical Laplace-Beltrami operator on a foliated geometry $N$
\begin{equation}
\label{eqn:FolLapBel}
\Delta =\frac{\partial^2 }{\partial \theta^2}+\frac{\partial^2 }{\partial \tau^2}.
\end{equation}
It is interesting to note that this restricted sum of squares operator is a horizontal Laplacian along the horizontal distribution $\mathcal{H}=\operatorname{span}\left\{\partial_\theta, \partial_\tau\right\} \subset T N$ .

The tropical inverse metric tensor $h$ can be constructed by either taking the direct limit of the inverse metric tensor under this Maslov dequantization or by noticing the tropical Laplacian whilst knowing that it still acts on a three-dimensional space parametrized by $(r,\theta,\tau)$. After a multiplicative renormalization to get rid of the divergent factor, one obtains
\begin{equation}
h \doteq\left[\begin{array}{ccc}
0 & 0 & 0 \\
0 & 1 & 0 \\
0 & 0 & 1
\end{array}\right] .
\end{equation}
The tropical metric tensor is also found to be
\begin{equation}
g \doteq\left[\begin{array}{ccc}
1 & 0 & 0 \\
0 & 0 & 0 \\
0 & 0 & 0
\end{array}\right] .
\end{equation}
Like in the case of tropicalized Riemann surfaces, they still satisfy the mutual invisibility law $gh=hg=0$.  

We now construct a square root of the tropical Laplacian by performing the same historical ansatz that Dirac made \cite{dirac1928quantum}
\begin{equation}
\slashed{D}=a \frac{\partial}{\partial r}+b \frac{\partial}{\partial \theta}+c \frac{\partial}{\partial \tau}.
\end{equation}
Here $a, b$, and $c$ are undetermined algebraic objects whose algebraic relations are determined by imposing the condition $\slashed{D}^2=\Delta$. Computing this square gives the following algebra
\begin{equation}
\begin{aligned}
&\{a, b\}=0, \quad\{b, c\}=0, \quad\{c, a\}=0, \\
& a^2=0, \quad b^2=1, \quad c^2=1,
\end{aligned}
\end{equation}
where, the anticommutator is $\{a,b\}=ab+ba$.  One can organize these in terms of tropical Clifford elements $\gamma^{I}=(a,b,c)$ and equivalently write down this algebra as
\begin{equation}
\left\{\gamma^I, \gamma^J\right\}=h^{IJ}. 
\end{equation}
We identify this as a degenerate Clifford algebra $C l(0,2,1)$, denoting that the signature associated with the quadratic form that generates the Clifford algebra has two positive non-degenerate directions and one null degenerate direction. 

We can construct a matrix representation for this degenerate tropical Clifford algebra by noticing that it contains two subalgebras. The first is a real nondegenerate 4-dimensional Clifford algebra $\mathcal{B}$  which is generated by $(1,b,c,bc)$. This algebra can be identified with the split quaternions. We now add a nilpotent generator $a$ which tells us that the tropical Clifford algebra is an 8-dimensional algebra $\mathcal{A}$ that decomposes as 
\begin{equation}
\mathcal{A}=\mathcal{B} \oplus a \mathcal{B}.
\end{equation}
This is reminiscent of the dual split quaternions where one adds a generator that lies in the center of the subalgebra $\mathcal{B}$. Instead, the nilpotent generator anticommutes with everything in $\mathcal{B}$.  We want to have an explicit matrix representation for the tropical Dirac operator. In order to find a minimal, irreducible, and faithful representation of the 8-dimensional tropical Clifford algebra, one can analyze the representation theory associated to the subalgebras. One finds that the minimum dimension of the representation over $\mathbb{R}$ is 4, preserving the dimension of the associated tropical spinor module. In this case, an explicit matrix realization is as follows. For the nilpotent tropical Clifford element, we have
\begin{equation}
a=\left(\begin{array}{cc}
0_2 & E \\
-E & 0_2
\end{array}\right)=\left(\begin{array}{cccc}
0 & 0 & 0 & 1 \\
0 & 0 & 0 & 0 \\
0 & -1 & 0 & 0 \\
0 & 0 & 0 & 0
\end{array}\right), \quad \text { with } \quad E=\left(\begin{array}{ll}
0 & 1 \\
0 & 0
\end{array}\right).
\end{equation}
For the other two Clifford elements, we have
\begin{equation}
\begin{gathered}
b=\left(\begin{array}{cc}
I_2 & 0 \\
0 & -I_2
\end{array}\right)=\left(\begin{array}{cccc}
1 & 0 & 0 & 0 \\
0 & 1 & 0 & 0 \\
0 & 0 & -1 & 0 \\
0 & 0 & 0 & -1
\end{array}\right), \quad 
c=\left(\begin{array}{cc}
0 & I_2 \\
I_2 & 0
\end{array}\right)=\left(\begin{array}{llcc}
0 & 0 & 1 & 0 \\
0 & 0 & 0 & 1 \\
1 & 0 & 0 & 0 \\
0 & 1 & 0 & 0
\end{array}\right) .
\end{gathered}
\end{equation}
We denote the 2x2 identity matrix as $I_2$. We call this representation, the purely real representation.

The corresponding tropical Dirac operator is then
\begin{equation}
\slashed{D}=\left(\begin{array}{cccc}
\partial_\theta & 0 & \partial_\tau & \partial_r \\
0 & \partial_\theta & 0 & \partial_\tau \\
\partial_\tau & -\partial_r & -\partial_\theta & 0 \\
0 & \partial_\tau & 0 & -\partial_\theta
\end{array}\right).
\end{equation}
As a check, we compute it's square and show that indeed it reproduces the tropical Laplace-Beltrami operator
\begin{equation}
\slashed{D}^2=\left[\begin{array}{cccc}
\partial_\theta^2+\partial_\tau^2 & 0 & 0 & 0 \\
0 & \partial_\theta^2+\partial_\tau^2 & 0 & 0 \\
0 & 0 & \partial_\theta^2+\partial_\tau^2 & 0 \\
0 & 0 & 0 & \partial_\theta^2+\partial_\tau^2
\end{array}\right].
\end{equation}
Interestingly, the tropical Dirac operator is sensitive to the radial direction unlike the tropical Laplace-Beltrami operator since it explicitly contains $\partial_r$ operators (cf. \eqref{eqn:FolLapBel}). 

One could pose the question of whether there is a complex representation of the degenerate Clifford algebra that allows us to work with smaller matrices than the 4x4 real representation we have constructed above. For our particular algebra, this turns out to be impossible. However, since we have a degenerate Clifford algebra, we have the additional option of working in the exterior algebra of the complex numbers $\bigwedge(\mathbb{C})$. To see this, we define the complex linear combinations for the subalgebra $\mathcal{B}$
\begin{equation}
\beta=\frac{1}{\sqrt{2}}(b+i c), \quad \beta^{\dagger}=\frac{1}{\sqrt{2}}(b-i c) .
\end{equation}
A short calculation then shows
\begin{equation}
\beta^2=\left(\beta^{\dagger}\right)^2=0, \quad\left\{\beta, \beta^{\dagger}\right\}=\beta \beta^{\dagger}+\beta^{\dagger} \beta=b^2+c^2=2.
\end{equation}
Hence, we can use the well-known fermionic oscillator matrix representation to write
\begin{equation}
\beta=\left(\begin{array}{cc}
0 & \sqrt{2} \\
0 & 0
\end{array}\right), \quad \beta^{\dagger}=\left(\begin{array}{cc}
0 & 0 \\
\sqrt{2} & 0
\end{array}\right).
\end{equation}
Inverting these to re-obtain the generators $b,c$, we get
\begin{equation}
\begin{aligned}
&b=\frac{1}{\sqrt{2}}\left(\beta+\beta^{\dagger}\right)=\frac{1}{\sqrt{2}}\left(\begin{array}{cc}
0 & \sqrt{2} \\
\sqrt{2} & 0
\end{array}\right)=\left(\begin{array}{ll}
0 & 1 \\
1 & 0
\end{array}\right),\\
&c=\frac{-i}{\sqrt{2}}\left(\beta-\beta^{\dagger}\right)=\frac{-i}{\sqrt{2}}\left(\begin{array}{cc}
0 & \sqrt{2} \\
-\sqrt{2} & 0
\end{array}\right)=\left(\begin{array}{cc}
0 & -i \\
i & 0
\end{array}\right)
.\end{aligned}
\end{equation}
We introduce a real Grassmann odd generator $\epsilon$, such that $\epsilon^2=0$, that allows us to represent the nilpotent generator $a$ as
\begin{equation}
\label{eqn:DegCliffA}
a=\epsilon\left(\begin{array}{cc}
1 & 0 \\
0 & -1
\end{array}\right).
\end{equation}
Notice that the matrix representations are just the conventional Pauli matrices up to a phase and a Grassmann parameter.  We interpret the addition of $\epsilon$ as supersymmetrizing the spinor bundle. However we do not have a supergeometry in the conventional sense since we do not supersymmetrize the foliated 3-manifold $N$.  In particular, we do not have superspace derivatives. We call this reduced matrix representation, the supersymmetric-complexified representation. The tropical Dirac operator then takes the form
\begin{equation}
\label{eqn:TropDiracCom}
\slashed{D}= \left(\begin{array}{ll}
\epsilon \partial_r & \partial_\theta-i \partial_\tau \\
\partial_\theta+i \partial_\tau & -\epsilon \partial_r
\end{array}\right).
\end{equation}
Indeed, this squares to the tropical Laplace-Beltrami operator acting on a complexified tropical spinor space.
\begin{equation}
\slashed{D}^2= \left(\begin{array}{ll}
\partial^2_\theta+\partial^2_\tau & 0\\
0& \partial^2_\theta+\partial^2_\tau
\end{array}\right).
\end{equation}

In correspondence with the fact that our method of tropicalization allows us to represent a real algebraic variety as a foliated complex geometry, it seems that our tropicalization forces us to work with real Clifford algebras if we want to avoid the introduction of Grassmann numbers to represent the nilpotent generators.  In the following table, we contrast the dimension of the conventional spin space $\hat{S}$ over a manifold $\hat{N}$ to it's tropical counterpart $S.$
\begin{equation}
\begin{array}{|c|c|c|}
\hline \operatorname{dim} \hat{N} & \operatorname{dim_\mathbb{C}} \hat{S}  & \operatorname{dim_{\mathbb{R}}} S\\
\hline 1 & 1 & *\\
2 & 2 & *\\
3 & 2 & 4\\
4 & 4 & *\\
5 & 4 & *\\
6 & 8 & *\\
7 & 8 & * \\
8 & 16 & * \\
9 & 16 & *\\
10 & 32 & * \\
\hline
\end{array}
\end{equation}
From the table, we see that it is not that the dimension of our original matrix representations in our 3-dimensional example has increased, but instead we had originally worked strictly over the reals. 

Before we further investigate  the properties of the tropical Dirac operator $D$, we study the arctic limit of the original Dirac operator and discuss what types of algebras arise. Performing the same arguments as before, one finds that the arctic Laplace-Beltrami operator is now 
\begin{equation}
\label{eqn:ArcLapBel}
\stackrel{\star}{\Delta}\phi=  \frac{\partial^2 \phi}{\partial r^2}.
\end{equation}
We use a star to denote arctic geometric objects. The arctic Clifford algebra now takes the following form
\begin{equation}
\begin{aligned}
& \{a, b\}=0, \quad\{b, c\}=0, \quad\{c, a\}=0, \\
& a^2=1, \quad b^2=0, \quad c^2=0.
\end{aligned}
\end{equation}
Organizing the arctic Clifford elements via the identification $\gamma^{I}=(a,b,c)$, we obtain
\begin{equation}
\left\{\gamma^I, \gamma^J\right\}=\stackrel{\star}{h}^{IJ} \doteq\left(\begin{array}{lll}
1 & 0 & 0 \\
0 & 0 & 0 \\
0 & 0 & 0
\end{array}\right).
\end{equation}
As a check, one can show that in the arctic limit, one obtains the above matrix representation for the arctic inverse metric tensor.

From our previous discussion, we can now identify the arctic Clifford algebra as $C l(0,1,2)$, denoting that we still have one nondegenerate positive direction but now two null directions associated with the two nilpotent generators $b,c$.  By the same arguments, one can construct a minimal dimensional representation of this algebra as
\begin{equation}
a=\left(\begin{array}{cccc}
1 & 0 & 0 & 0 \\
0 & -1 & 0 & 0 \\
0 & 0 & -1 & 0 \\
0 & 0 & 0 & 1
\end{array}\right), \quad b=\left(\begin{array}{cccc}
0 & 0 & 0 & 0 \\
1 & 0 & 0 & 0 \\
0 & 0 & 0 & 0 \\
0 & 0 & 1 & 0
\end{array}\right), \quad c=\left(\begin{array}{cccc}
0 & 0 & 0 & 0 \\
0 & 0 & 0 & 0 \\
1 & 0 & 0 & 0 \\
0 & -1 & 0 & 0
\end{array}\right).
\end{equation}
The arctic Dirac operator $\stackrel{\star}{\slashed{D}}$ then takes the form
\begin{equation}
\stackrel{\star}{\slashed{D}}=\left(\begin{array}{cccc}
\partial_r & 0 & 0 & 0 \\
\partial_\theta & -\partial_r & 0 & 0 \\
\partial_{\tau} & 0 & -\partial_r & 0 \\
0 & -\partial_\tau & \partial_\theta & \partial_r
\end{array}\right).
\end{equation}
One can check that indeed, the square of the arctic Dirac operator gives the arctic Laplace-Beltrami operator $\stackrel{\star}{\Delta}$ \eqref{eqn:ArcLapBel}
\begin{equation}
\stackrel{\star}{\slashed{D}}^2=\left[\begin{array}{cccc}
\partial_r^2 & 0 & 0 & 0 \\
0 & \partial_r^2 & 0 & 0 \\
0 & 0 & \partial_r^2 & 0 \\
0 & 0 & 0 & \partial_r^2
\end{array}\right].
\end{equation}

If one wishes to reduce down the dimension of the matrix representation chosen, we are once again allowed to introduce Grassmann numbers to represent the nilpotent generators $b,c$.  One might expect that we have an extended supersymmetry in this case i.e., one Grassmann odd generator per collapsed direction. However, the degenerate Clifford algebra forces us to work with a single Grassmann number $\epsilon$, in order to satisfy the anticommutativity relation between the two nilpotent generators. The explicit representations are
\begin{equation}
a=\left(\begin{array}{cc}
1 & 0 \\
0 & -1
\end{array}\right), \quad b=\left(\begin{array}{cc}
0 & \epsilon \\
0 & 0
\end{array}\right), \quad c=\left(\begin{array}{cc}
0 & 0 \\
\epsilon & 0
\end{array}\right).
\end{equation}
The arctic Dirac equation is then
\begin{equation}
\stackrel{\star}{\slashed{D}}=\left[\begin{array}{cc}
\partial_r & \epsilon \partial_\theta \\
\epsilon \partial_\tau & -\partial_r
\end{array}\right].
\end{equation}
Notice that this representation does not introduce a complex number. We call this a supersymmetric real representation of the degenerate Clifford algebra.

\section{Applications of Tropical Dirac Operators}
\label{sec:PropTropDir}
In this section, we investigate the solutions associated to the tropical Dirac equation for the supersymmetric-complexified representation. Using the representations of the previous section, we write down a simple tropical spinor Lagrangian and write down the essential elements required to canonically quantize the theory and/or build its path integral formulation. We then twist these tropical Dirac operators by a vector bundle to derive the tropical analog of the well-known Lichnerowicz identity. We find that this defines the appropriate notion of tropical Yang-Mills curvature. 

We define the tropical spinor space as the representation space that the tropical Clifford algebra acts on. An element of this representation space is what we call a tropical or arctic spinor, depending on the limit we took. We begin by naturally defining the massless tropical Dirac equation as
\begin{equation}
   \slashed{D} \Psi=0.
\end{equation}
Here, $\Psi$ is a tropical spinor and $\slashed{D}$ is the tropical Dirac operator given by \eqref{eqn:TropDiracCom}. 

We begin by working in the supersymmetric-complexified representation. Consequently, tropical spinors $\psi$ are represented by a doublet of complex spinors that has a decomposition into a Grassmann degree 0 and a Grassmann degree 1 component. 
\begin{align}
\Psi=\begin{pmatrix}\psi_1 \\ \psi_2\end{pmatrix}+\epsilon\begin{pmatrix}
    \Phi_1 \\ \Phi_2\end{pmatrix}.
\end{align}
Both $\psi_{\mu}$ and $\Phi_{\mu}$, with $\mu=\{1,2\}$, further decompose into real and imaginary parts. The massless tropical Dirac equation is then
\begin{equation}
\begin{aligned}
& \left(\partial_\theta-i \partial_{\tau}\right) \psi_2+\epsilon\left(\partial_r \psi_1+\left(\partial_\theta-i \partial_{\tau}\right) \Phi_2\right)=0, \\
& \left(\partial_\theta+i \partial_{\tau}\right) \psi_1+\epsilon\left(\left(\partial_\theta+i \partial_{\tau}\right) \Phi_1-\partial_r \psi_2\right)=0.
\end{aligned}
\end{equation}
By the linear independence of the Grassmann number $\epsilon$, these can be further decomposed as
\begin{equation}
\begin{aligned}
& \left(\partial_\theta+i \partial_\tau\right) \psi_1=0, \\
& \left(\partial_\theta-i \partial_\tau\right) \psi_2=0, \\
& \partial_r \psi_1+\left(\partial_\theta-i \partial_{\tau}\right) \Phi_2=0, \\
& \left(\partial_\theta+i \partial_{\tau}\right) \Phi_1-\partial_r \psi_2=0.
\end{aligned}
\end{equation}
The solutions to the massless tropical spinor equation can be probed by first solving the standard Dolbeault-Dirac equations in $(r,\theta)$  for $\psi_{\mu}$. Once these solutions are constructed, they can be used as a forcing function for the equations that evolve $\Phi_{\mu}$.  The leftover equations are then of the form
\begin{equation}
\begin{aligned}
& \left(\partial_\theta-i \partial_{\tau}\right) \Phi_2=-\partial_r \psi_1, \\
& \left(\partial_\theta+i \partial_{\tau}\right) \Phi_1=\partial_r \psi_2,
\end{aligned}
\end{equation}
which can then be solved by the Greens functions associated to the standard Dolbeault-Dirac operator along the $(\theta,\tau)$ directions. This demonstrates that the additional spinor fields $(\Phi_1,\Phi_2)$ are determined by the spinor fields ($\psi_1,\psi_2$). These equations are suggestive of an additional gauge symmetry. In fact, one can directly see that the equations of motion for the spinor $\Phi_\mu$, admit a gauged shift symmetry where they are deformable by any arbitrary function of the radial direction $\alpha(r)$.

In order to construct a tropical Dirac Lagrangian, we define a formal Dirac conjugate as
\begin{equation}
\Psi^{\dagger}=\left(\psi_1^{\dagger}-\epsilon \Phi_1^{\dagger}, \psi_2^{\dagger}-\epsilon \Phi_2^{\dagger}\right).
\end{equation}
The dagger refers to the complex conjugation of complex numbers.  We define the conjugation of the Grassmann number $q$ in this way, since the real-valued norm of a Grassmann number can be canonically defined as
\begin{equation}
\|q\|^2=(x-\epsilon y)(x+\epsilon y)=x^2.
\end{equation}
The upshot of this is that we introduce a new global symmetry for our Lagrangians by performing a rotation with a Grassmann number with a real symmetry parameter $\Bucket$. As an example, consider a Grassmann number valued real scalar $\phi$ that has been transformed by this nilpotent symmetry transformation
\begin{equation}
\phi^{\prime}=e^{\Bucket  \epsilon} \phi=(1+\Bucket \epsilon) \phi=\phi+\Bucket \epsilon \phi
\end{equation}
Under this symmetry transform, the norm of a Grassmann number remains invariant
\begin{equation}
\left\|\phi^{\prime}\right\|^2=(\phi-\Bucket \epsilon \phi)(\phi+\Bucket\epsilon \phi)=\|\phi\|^2.
\end{equation}
We call this global symmetry, the bucket symmetry. 

The simplest candidate for a Lagrangian density is
\begin{equation}
    \mathcal{L}=\Psi^{\dagger}\slashed{D}\Psi
\end{equation}
Now, we proceed with a Dirac-Bergmann canonical quantization of the massless tropical Dirac Lagrangian. The Lagrangian density reduces down to
\begin{align}
    \mathcal{L}&=\psi_1^\dagger\left(\left(\partial_\theta-i \partial_{\tau}\right) \psi_2+\epsilon\left(\partial_r \psi_1+\left(\partial_\theta-i \partial_{\tau}\right) \Phi_2\right)\right) -\epsilon \Phi_1^\dagger\left(\partial_\theta-i \partial_{\tau}\right) \psi_2 \\ \nonumber
    &+\psi_2^\dagger\left(\left(\partial_\theta+i \partial_{\tau}\right) \psi_1+\epsilon\left(\left(\partial_\theta+i \partial_{\tau}\right) \Phi_1-\partial_r \psi_2\right)\right) 
    -\epsilon \Phi_2^\dagger\left(\partial_\theta+i \partial_{\tau}\right) \psi_1.
\end{align}
The conjugate momenta for the spinor components $(\psi_1,\psi_2, \Phi_1,\Phi_2)$ are
\begin{align}
    \Pi_{\psi_1}&=i\psi_2^\dagger-i\epsilon \Phi_2^\dagger, \quad \; \, \Pi_{\Phi_1}=i\epsilon \psi_2^\dagger, \\
    \Pi_{\psi_2}&=-i\psi_1^\dagger+i\epsilon \Phi_1^\dagger, \quad  \Pi_{\Phi_2}=-i\epsilon\psi_1^\dagger,
\end{align}
while the conjugate momenta of the components $(\psi_1^\dagger,\psi_2^\dagger,\Phi_1^\dagger,\Phi_2^\dagger)$ are
\begin{align}
    \Pi_{\psi_1^\dagger}=\Pi_{\psi_2^\dagger}=\Pi_{\Phi_1^\dagger}=\Pi_{\Phi_2^\dagger}= 0.
\end{align}
Therefore, we have eight primary constraints that do not lead to any further constraints.  To proceed with the quantization, we need to classify the class of constraints that we have. One can check that the matrix of constraints vanishes due to the Grassmann number $\epsilon$ and, hence, we can expect that some of constraints are first class. We treat Poisson brackets of constraints whose $\epsilon$ independent part vanishes as first class.
These first class constraints generate gauge symmetries
\begin{align}
    \delta \Phi_1=\alpha_{\Phi_1}, \quad \delta \Phi_2=\alpha_{\Phi_2}, \quad \delta \Phi_1^\dagger=\alpha_{\Phi_1^\dagger}, \quad \delta \Phi_2^\dagger=\alpha_{\Phi_2^\dagger},
\end{align}
and second class constraints modify the standard Dirac bracket. We remove the first class constraints by performing gauge fixing where we set the additional spinor fields to vanish
\begin{align}
\label{eqn:GaugeFix}
    \Phi_1=\Phi_2=\Phi_1^\dagger=\Phi_2^\dagger=0.
\end{align}
One can check that this gauge fixing condition is consistent. In particular, their standard Dirac brackets with the first class constraints does not vanish. Once we impose the gauge fixing conditions, we can canonically quantize using the modified Dirac bracket. We impose the canonical anti-commutation relations on the leftover fields
\begin{align}
    \{\psi_1,\Pi_{\psi_1}\}=\{\psi_2,\Pi_{\psi_2}\}=i.
\end{align}
We denote the constraints as 
\begin{align}
    \chi_1&=\Pi_{\psi_1}-i\psi_2^\dagger\approx 0, \quad \chi_3=\Pi_{\psi_1^\dagger}\approx 0, \\
    \chi_2&=\Pi_{\psi_2}+i\psi_1^\dagger\approx 0, \quad \chi_4=\Pi_{\psi_2^\dagger}\approx 0, 
\end{align}
together with $4$ gauge fixing conditions that set \eqref{eqn:GaugeFix}. The modified Dirac brackets is then
\begin{align}
    \{\psi_i,\psi_j^\dagger\}_{D}=\{\psi_i,\psi_j^\dagger\}-\{\psi_i,\chi_k\}C^{kl}\{\chi_l,\psi_j^\dagger\},
\end{align}
where $C$ is the inverse of the constraints matrix after gauge fixing. In particular, we notice that
\begin{align}
    \{\psi_1,\psi_1^\dagger\}_D=-\{\psi_2,\psi_2^\dagger\}_D=1,
\end{align}
with all other anti-commutators vanishing. In order to construct the full Hilbert space, we need to have explicit solutions for the tropical Dirac equation that obey the matching conditions of tropical graphs. We leave this for future work in an upcoming paper \cite{WIP1}.

In the case where $\Phi_{\mu}$ and its conjugate vanish, as in \eqref{eqn:GaugeFix}, these have a simple interpretation. Our tropical spinor fields must be constant and transverse to the leaves of the foliation. In the context of tropical graphs, this means that they must have a frozen radial value until they reach the vertex point where the additional matching conditions must be imposed. The origin of this condition comes from the nilpotent generator $a$. The other two equations of motion are the standard Dolbeault-Dirac operators. 

Now, we twist our tropical Dirac operator by a vector bundle $E$ defined over $N$. In order to do so, we apply the minimal coupling prescription. Associated to a U$(1)$ gauge symmetry group generated by $\beta(r,\theta,\tau)$, our twisted tropical spinors transform as
\begin{equation}
\Psi^{\prime}(r, \theta, \tau)=e^{i \beta(r, \theta, \tau)} \Psi(r, \theta, \tau).
\end{equation}
This introduces a gauge field $A$ on the foliated manifold $N$, which defines the twisted tropical Dirac operator 
\begin{equation}
\begin{aligned}
\slashed{D}^A & =\gamma^I D_I^A \\
& =\gamma^I\left(\partial_I+A_I\right) \\
& =a\left(\partial_r+A_r\right)+\gamma^i\left(\partial_i+A_i\right),
\end{aligned}
\end{equation}
where we use uppercase index to denote directions $(r,\theta,\tau)$ and the lowercase index to denote directions $(\theta,\tau)$.
As in \eqref{eqn:DegCliffA}, $a$ is our nilpotent generator associated to a Grassmann number in the supersymmetric-complexified representation. $\gamma^{i}=(b,c)$ refers to the Clifford elements along the $(\theta,\tau)$ directions.  The explicit matrix representation for the twisted tropical Dirac operator is
\begin{equation}
\slashed{D}^A=\left[\begin{array}{ll}
\epsilon\left(\partial_r+A_r\right) & \;\; \left(\partial_\theta-i \partial_{\tau}\right)+\left(A_\theta-i A_\tau\right) \\
\left(\partial_\theta+i \partial_{\tau}\right)+\left(A_\theta+i A_\tau\right) & \;\; -\epsilon\left(\partial_r+A_r\right).
\end{array}\right]
\end{equation}
Using either the matrix representation or the abstract algebraic relations, we can compute the tropical Lichnerowicz identity
\begin{equation}
\left(\slashed D^A\right)^2={\gamma}^I {\gamma}^J\left(\partial_I+A_I\right)\left(\partial_J+A_J\right).
\end{equation}
Breaking both pieces into symmetric and antisymmetric components, one finds
\begin{align}
    (\slashed{D}^A)^2&=h^{I J} D_I^A D_J^A+\frac{1}{2}\left[{\gamma}^I, {\gamma}^J\right]\left(\partial_I A_J-\partial_J A_I\right) \\
& =h^{i j} D_i^A D_j^A+\frac{1}{2}\left[{\gamma}^I, {\gamma}^J\right]\left(\partial_I A_J-\partial_J A_I\right).
\end{align}
Notice that the symmetric portion reduces down to the tropical inverse metric tensor and, hence, we only pick out derivatives along the $(\theta,\tau)$ directions. The upshot of the Lichnerowicz identity is that it tells us how a tropical Yang-Mills curvature should appear. As usual, we define the components of the tropical Yang-Mills curvature as 
\begin{equation}
    F_{IJ}=\partial_IA_J-\partial_JA_I.
\end{equation}
The antisymmetric contribution can be simplified further by using the Grassmann-complexified matrix representations. The relevant commutators evaluate to
\begin{equation}
\begin{aligned}
& {[a, b]=a b-b a=2 i \epsilon\left(\begin{array}{cc}
0 & -i \\
i & 0
\end{array}\right)}, \\
& {[a, c]=a c-c a=-2 i \epsilon\left(\begin{array}{ll}
0 & 1 \\
1 & 0
\end{array}\right)}, \\
& {[b, c]=b c-c b=2 i\left(\begin{array}{cc}
1 & 0 \\
0 & -1
\end{array}\right)}.
\end{aligned}
\end{equation}
The tropical Lichnerowicz identity then takes the form
\begin{equation}
(\slashed{D}^{A})^2=\left[\begin{array}{ll}
\left(\partial_i+A_i\right)^2+i F_{\theta \tau} & \;\; \epsilon F_{r \theta}-i \epsilon F_{r \tau} \\
-\epsilon F_{r \theta}-i \epsilon F_{r \tau} & \;\; \left(\partial_i+A_i\right)^2-i F_{\theta \tau}
\end{array}\right].
\end{equation}
Notice that the curvature terms that are related to the foliation direction $r$ come along with a Grassmann odd element $\epsilon$.  This will be an essential observation in calculating path integrals associated to these tropical Dirac operators. 

 A simple U$(1)$ gauge invariant Lagrangian density can then be written down as
\begin{equation}
\mathcal{L}=\Psi^{\dagger} \slashed{D}^A \Psi.
\end{equation}
We can expand this into components to obtain a Lagrangian density that decomposes into a Grassmann degree zero term $\mathcal{L}_0$ and a Grassmann degree one term $\mathcal{L}_1$ 
\begin{equation}
\mathcal{L}_0=\psi_1^{\dagger}\left(\left(\partial_\theta-i \partial_\tau\right)+\left(A_\theta-i A_\tau\right)\right) \psi_2+\psi_2^{\dagger}\left(\left(\partial_\theta+i \partial_\tau\right)+\left(A_\theta+i A_\tau\right)\right) \psi_1.
\end{equation}
Notice that the Grassmann degree zero component is just the standard Dirac-Dolbeault Lagrangian density along the $(\theta,\tau)$ directions. The Grassmann degree one term is 
\begin{align}
\mathcal{L}_1&=\epsilon\Big[\psi_1^{\dagger}\left(\partial_r+A_r\right) \psi_1-\psi_2^{\dagger}\left(\partial_r+A_r\right) \psi_2  \\ \nonumber
&-\Phi_2^{\dagger}\left(\left(\partial_\theta+i \partial_\tau\right)+\left(A_\theta+i A_\tau\right)\right) \psi_1+\psi_1^{\dagger}\left(\left(\partial_\theta-i \partial_\tau\right)+\left(A_\theta-i A_\tau\right)\right) \Phi_2\\ \nonumber
&+\psi_2^{\dagger}\left(\left(\partial_\theta+i \partial_\tau\right)+\left(A_\theta+i A_\tau\right)\right) \Phi_1 -\Phi_1^{\dagger}\left(\left(\partial_\theta+i \partial_\tau\right)+\left(A_\theta+i A_\tau\right)\right) \psi_2\Big].
\end{align}

\section{Tropical Localization and Schwinger Effect}
\label{sec:TropLoc}

The consequences of introducing a Grassmann number into the tropical spinor bundle is not clear at this moment. It is suggestive that the corresponding path integral should now have an additional integration over the Grassmann number. As a zero-dimensional schematic, we expect our path integrals to be formally evaluated as follows
\begin{equation}
\int d \epsilon e^{-S_0-\epsilon S_1}=\int d \epsilon e^{-S_0}\left(1-\epsilon S_1\right)=-e^{-S_0} S_1.
\end{equation}
This is reminiscent to how full-fledged supersymmetric localization \cite{ewtsm, Nekrasov:2002qd} occurs in zero-dimensional supersymmetric path integrals. However, in our case, we do not have supergeometry that extends the foliated manifold. Instead, we have a supersymmetry strictly on the spinor bundle. If we were to couple these tropical spinors to any other field, the other fields need not satisfy any additional supersymmetry. We will eventually show how one can utilize this additional fermionic localization to perform path integral calculations in \cite{WIP2}. 

Consequently, it appears that we have an example of localization where all our fields are strictly fermionic. In fact, one can trace this localization back to an additional topological fermionic symmetry that localizes the appropriate integrals. In order to make the distinction from standard supersymmetric localization, we call this new class of localizations, \textit{tropical localization} and it's arctic cousin, \textit{arctic localization}. Recently, it has been shown in \cite{Choi:2025jis}, based on \cite{Choi:2021yuz} that hidden fermionic symmetries can be used to localize worldline calculations in much the same way as is suggested in this paper. Our reasoning for this claim is the following. The original motivation for investigating the tropicalization of geometric objects was significantly motivated from trying to probe the Schwinger-Keldysh wedge region \cite{ssk, neq, keq} which we have argued in previous papers \cite{trsm, Albrychiewicz:2024tqe} is supposed to correspond to a tropicalized field theory that describes the asymptotic region of the moduli space of the string worldsheet. We expect the zero-dimensional fermionic symmetry which the authors of \cite{Choi:2025jis} arises from the arctic limit of the moduli space of worldline moduli $T$.  One can expect this to be the case since a short calculation demonstrates the double analytic continuation of the worldline on the $+$ branch of the SK contour and the $-$ branch localizes onto the singular points of the moduli space. This can be seen explicitly for the Schwinger effect \cite{Schwinger:1951nm} as shown below.

For the derivation of the Schwinger effect for bosons pair production using the worldline formalism \cite{Affleck:1981bma, Kim:2000un, Dunne:2005sx}, we start with the standard Euclidean path integral on a target space $\mathcal{M}=\mathbb{R}^4$ with a constant background field in the Fock Schwinger gauge $A_\mu=\frac{F_{\mu \nu}}{2} X^{\nu}$. The standard accepted calculation for the Schwinger effect begins with a loop space integral which by direct calculation reduces down to the space of constant loops after the fluctations are integrated out. This introduces an additional symmetry given by the diffeomorphisms of the circle that need to be gauged fixed. In particular, the BRST ghost sector brings in a factor of $1/T$, where $T$ is the modulus of the worldline after gauge fixing the worldline diffeomorphisms such that the local expression for the einbein is $e(\tau)=2$ to match the literature.  The result is simply 
\begin{equation}
Z= \int_0^{\infty} \frac{d T}{T} \frac{1}{(4 \pi T)^{D / 2}} \frac{q F T}{\sin q F T} \operatorname{Vol}\mathcal{M} e^{-m^2 T},
\end{equation}
where $\operatorname{Vol}\mathcal{M}$ arises from the zero mode integration. The new method is now implemented by noticing that the integrand has poles for $T_n=\frac{n \pi}{q F}$ and thus we propose a new integration cycle $\Gamma_+$. One avoids these poles by deforming the integration cycle by semicircles while retaining the horizontal contribution in between them. The arcticalization procedure is implemented by parameterizing the moduli space integration cycle by 
\begin{equation}
    \hat{T}=e^{\hbar r + i\theta}.
\end{equation}

Inserting this into the moduli space integral and taking the limit $\hbar \rightarrow 0$, reduces this down to an integral over the angular components of the deformed contour $\Gamma_+$, giving us precisely the imaginary contributions of the worldline instantons. These appear as half residues with clockwise orientation in both cases. This corresponds to the forward time branch of the Schwinger Keldysh contour as shown in \figref{fig:ArcContour}. The result after performing the above procedure is
\begin{equation}
\operatorname{Arctic} Z_{+}=\frac{-q^2 F^2}{16 \pi^3} \sum_{n=1}^{\infty} \frac{(-1)^n}{n^2} e^{-\frac{m^2 n \pi}{q F}} \operatorname{vol}\mathcal{M} \text{.}
\end{equation} 
\begin{figure}[ht]
\centering
\resizebox{\textwidth}{!}{
\begin{tikzpicture}[
    scale=1.5,                 
    thick,                      
    pole/.style={black, thick}, 
    midarrow/.style={postaction={decorate, decoration={
        markings, mark=at position 0.55 with {\arrow[scale=1.5]{>}}
    }}}
]
    \draw[->] (0, -0.5) -- (0, 2) node[above] {$\text{Im}(T)$}; 
    \draw[-] (-1, 0) -- (9.5, 0) node[right] {$\text{Re}(T)$}; 
    \foreach \x in {0, 2, 4, 6, 8} {
        \draw[pole] (\x,0) ++(-0.15, -0.15) -- ++(0.3, 0.3);
        \draw[pole] (\x,0) ++(-0.15, 0.15) -- ++(0.3, -0.3);
        \draw[color={rgb, 255:red, 74; green, 144; blue, 226}] 
            (\x + 0.4, 0) arc (0:180:0.4);
        \ifnum\x<8
            \draw[color={rgb, 255:red, 74; green, 144; blue, 226}, midarrow] 
                (\x + 0.4, 0) -- (\x + 2 - 0.4, 0);
        \else
            \draw[color={rgb, 255:red, 74; green, 144; blue, 226}, midarrow] 
                (\x + 0.4, 0) -- (\x + 1.0, 0);
        \fi
    }
\end{tikzpicture}
}
\caption{The contour used for $\Gamma+$. Notice that as $\hbar$ goes to zero the horizontal real part no longer contributes to the integral, however, the integration cycle itself fully extends to the origin as expected. There is a similar contour for the backwards branch whose contribution to the final sum is equal.}
\label{fig:ArcContour}
\end{figure}

This is not what we would get from the Schwinger effect, it is off by a $1/2$ factor; but recall that we need to add the backwards branch $\Gamma_-$\footnote{This is another piece of evidence suggesting that this is a new method of calculation because the usual procedure with worldsheet instantons does not require an additional contribution from the backwards branch. Here, the poles on the negative branch correspond to worldsheet "anti-instantons".}.  We add this by performing the same Euclidean path integral but with einbein being gauged fixed to be negative. For this backwards time branch, we deform the integration cycle over the moduli space once again by evading the poles. Now we arcticalize each leftover integral before we sum them. Summing up both contributions gives precisely the Schwinger effect. There is no pole that contributes at $T=0$. For ease of comparison, we evaluate this when the target spacetime dimension is $4$. The final result is  
\begin{equation}
\operatorname{Arctic} Z_{+} + \operatorname{Arctic} Z_{-}=\frac{-q^2 F^2}{8 \pi^3} \sum_{n=1}^{\infty} \frac{(-1)^n}{n^2} e^{-\frac{m^2 n \pi}{q F}} \operatorname{vol}\mathcal{M} \text{.}
\end{equation} 
It is interesting to compare this to standard methods of obtaining the Schwinger effect by taking a logarithm after the path integral is done and then explicitly extracting the decay rate through the imaginary part of the effective action. Thus, we have evidence that the way our arcticalization procedure works is via implementing the logarithm within the integral instead of after the Euclidean path integral. Mathematically, this is stated as
\begin{equation}
\operatorname{Im} \Gamma =-\left(\operatorname{Arctic} Z_{+}+\operatorname{Arctic} Z_{-}\right) \text{.}
\end{equation}

There is one additional potential interesting application for arctic localization. Consider the case of an infinitely long cylindrical surface $\hat\Sigma$. In the arctic limit, the leaves of the foliation run along the radial direction, effectively collapsing this direction and leaving behind a circle $S^1$. Intuitively, this allows us to probe how the Dirac operator behaves asymptotically at infinite radius. The question of how the Dirac operator behaves on noncompact manifolds has been studied before in the context of the Atiyah-Patodi-Singer index theorem \cite{atiyah1975spectral} where one is interested in extensions of the Atiyah-Singer index theorem to the case of a manifold with boundaries. The APS index theorem has been used to probe condensed matter physics on the surface of materials in topological phases \cite{fukayaAPS}, and as well in the analysis of anomaly inflow \cite{wittenInflow}. 

As explained in \cite{melroseAPS}, one is generically able to construct rigorous proofs of the APS index theorem by either directly dealing with the boundary, or by extending the manifold to a noncompact setting where the boundary is described by an infinite cylinder. In this asymptotic limit, we expect that the geometry transverse to the radial direction is what gives rise to a spectral invariant associated to a self-adjoint operator $A$ known as the eta invariant which can be defined by a regularization
\begin{equation}
\eta_A(s)=\sum_{\lambda \neq 0} \frac{\operatorname{sign}(\lambda)}{|\lambda|^s} = \frac{1}{\Gamma\left(\frac{s+1}{2}\right)} \int_0^{\infty} t^{\frac{s-1}{2}} \operatorname{Tr}\left(A e^{-t A^2}\right) d t,
\end{equation}
such that the eta invariant is given by $\eta_A(0)$. The eta invariant measures the spectral asymmetry between the eigenvalues $\lambda$ of the self-adjoint operator $A$ by counting the difference of the number of positive eigenvalues and the number of negative eigenvalues.  For the case of a foliated geometry given by the arctic limit, we can expect that the path integral is effectively given by the eta invariant. We will leave this as an open question. 

\section{Conclusions and Outlook}
\label{sec:Conclusions}

In this paper, we constructed a tropicalized Laplacian and it's associated tropical Dirac operator through the Maslov dequantization procedure. We found that one of our examples of degenerated Laplacians is a horizontal Laplacian, it is tempting to connect this to the hypo-elliptic sub-Laplacians which are studied in the context of nonholonomic Riemannian geometry/sub-Riemannian geometry \cite{bellaiche1996sub}. However, the vector fields that generate our distributions are not bracket generating so we do not have hypo-ellipticity. It would be interesting to see if one is able to take a more geometric approach to tropicalization that would result in a hypo-elliptic differential operator \cite{hormander1961hypoelliptic} such as the the sub-Laplacian since then hypo-ellipticity guarantees the existence of a heat kernel which can explicitly calculated via stochastic analysis \cite{calin2023stochastic}. 

We also found an alternative tropical limit of the Laplacian which was non-linear in nature but yielded non-trivial tropical curves. It would be interesting to explore whether this differential operator nonetheless admits a square root despite its nonlinearity and also to investigate what sort of novel algebraic structures could appear.  If such a nonlinear square root exists, in order to not get confused with the tropical spinors introduced in this paper, we call the corresponding nonlinear spinoral objects, \textit{spinolinears}.

We investigated the tropical Dirac operator in the flat 3-dimensional case in order to more explicitly see what new structures appear. The tropical Clifford algebra was obtained in two distinct ways; the first by constructing an explicit square root of the tropical Laplacian, and the second by taking the direct tropical limit of the Clifford algebra. Both methods led to a degenerate Clifford algebra which contains nilpotent generators associated to the tropicalized directions. The representation space of the degenerate Clifford algebra defined the tropical spinors. We showed that there exists another limit associated to the $\hbar \rightarrow \infty$ limit that leads to the notion of arctic spinors. In trying to construct complex representations of the tropical Clifford algebra, one is forced to introduce a Grassmann number that supersymmetrizes the spinor bundle. One could in principle study further properties of these degenerate tropical Clifford algebras and see if there are analogous concepts that can be defined such as tropical chirality and the analogs of different spinor conditions like the Majorana condition.

The Grassmann number naively doubles the number of fields that are present in the theory since spinors now become Grassmann number-valued. However, upon implementing Dirac-Bergmann's quantization procedure on a simple tropical Dirac Lagrangian, we find that we have a combination of first and second class constraints that allows us to eliminate these fields and show that the original spinors are constant transverse to the leaves of the foliation.  In addition to this, we find that our tropical Dirac Lagrangian enjoys a new global nilpotent symmetry generated by the Grassmann number that we call the bucket symmetry. It would be interesting to see if this bucket symmetry can be gauged and then coupled to any conserved symmetry charges that it generates. 

Although we have worked in the zero curvature case for simplicity, an immediate extension of these story can be constructed by trying to generalize this for specific Riemannian geometries that can be deformed in a similar way. In order to do this, one first needs to learn how to properly deal with Riemann curvature terms in the tropical limit. We provide a solution to this issue in \cite{WIP2}.

In particular, one would like to push the Maslov dequantization that works in symplectic geometry case \cite{Albrychiewicz:2024tqe, Albrychiewicz:2025hzt} to the odd-dimensional case associated to contact geometries and the higher order Jordan structures predicted in \cite{trsm}. In order to extract potentially useful invariants of the aforementioned geometries, one needs to figure out how to properly supersymmetrize these degenerate field theories in order to construct the tropical analog of topological field theories in the target space. 

We have shown that we can minimally couple these tropical spinors to gauge fields, and by the Lichnerowicz identity, we identify how a tropical notion of Yang-Mills curvature should arrive. In particular, the curvature terms along the foliation come with a nilpotent factor of the Grassmann number.  Finally, we discussed how this additional supersymmetry that appears in the spinor bundle can potentially be used to localize calculations in these tropical path integrals. We will present the complete details of this path integral in an upcoming paper on the path integral quantization of the tropical Dirac equation \cite{WIP1, WIP2}.  Although, we have primarily investigated the tropical limit in this paper, the arctic limit is a distinct limit since the original foliation direction is swapped with the direction that parametrizes the leaves of the foliation. It would be interesting to repeat the constructions in this paper for the arctic limit and investigate whether or not there are any novelties for the arctic theory or dualities between the tropical and arctic theories.

\subsection*{Acknowledgements}
We gratefully acknowledge Petr Ho\v{r}ava for his foundational contributions and insightful guidance in the development of the program of tropological sigma models. We would like to also acknowledge Ori Ganor and Viola Zixin Zhao for useful discussions and comments on this manuscript. This work has been supported by the Leinweber Institute for Theoretical Physics.

\bibliographystyle{JHEP}
\bibliography{Bibliography/0BTropicalDirac}

\end{document}